\documentclass[twocolumn,trackchanges]{aastex631}

\usepackage{multirow}

\defcitealias{poggianti+2017b}{P17b}

\shorttitle{Exploring the AGN-ram pressure stripping connection}
\shortauthors{Peluso et al.}
\graphicspath{{./}{figures/}}

\begin{document}

\title{Exploring the AGN-ram pressure stripping connection in local clusters}

\author[0000-0001-5766-7154]{Giorgia Peluso}
\affiliation{INAF-Padova Astronomical Observatory, Vicolo dell’Osservatorio 5, I-35122 Padova, Italy}
\affiliation{Dipartimento di Fisica e Astronomia "G. Galilei", Università di Padova, Vicolo dell’Osservatorio 3, 35122 Padova, Italy
}
\author[0000-0003-0980-1499]{Benedetta Vulcani}\affiliation{INAF-Padova Astronomical Observatory, Vicolo dell’Osservatorio 5, I-35122 Padova, Italy}

\author[0000-0001-8751-8360]{Bianca M. Poggianti}
\affiliation{INAF-Padova Astronomical Observatory, Vicolo dell’Osservatorio 5, I-35122 Padova, Italy}

\author[0000-0002-1688-482X]{Alessia Moretti}\affiliation{INAF-Padova Astronomical Observatory, Vicolo dell’Osservatorio 5, I-35122 Padova, Italy}

\author[0000-0002-3585-866X]{Mario Radovich}\affiliation{INAF-Padova Astronomical Observatory, Vicolo dell’Osservatorio 5, I-35122 Padova, Italy}

\author[0000-0001-5303-6830]{Rory Smith}\affiliation{Korea Astronomy and Space Science Institute (KASI), 776 Daedeokdae-ro, Yuseong-gu, Daejeon 34055, Republic of Korea}

\author[0000-0003-2150-1130]{Yara L. Jaff\'e}\affiliation{Instituto de F\'isica y Astronom\'ia, Facultad de Ciencias, Universidad de Valpara\'iso, Avda. Gran Breta\~na 1111, Casilla 5030, Valpara\'iso, Chile}

\author{Jacob Crossett}\affiliation{Instituto de F\'isica y Astronom\'ia, Facultad de Ciencias, Universidad de Valpara\'iso, Avda. Gran Breta\~na 1111, Casilla 5030, Valpara\'iso, Chile}

\author[0000-0002-7296-9780]{Marco Gullieuszik}\affiliation{INAF-Padova Astronomical Observatory, Vicolo dell’Osservatorio 5, I-35122 Padova, Italy}

\author[0000-0002-7042-1965]{Jacopo Fritz}\affiliation{Instituto de Radioastronomía y Astrofísica, UNAM, Campus Morelia, A.P. 3-72, C.P. 58089, Mexico}

\author[0000-0003-1581-0092]{Alessandro Ignesti}\affiliation{INAF-Padova Astronomical Observatory, Vicolo dell’Osservatorio 5, I-35122 Padova, Italy}









\begin{abstract}
Ram-pressure stripping by the intracluster medium (ICM) is one of the most advocated mechanisms that affect the properties of cluster galaxies. A recent study  based on a small sample has found that many galaxies showing strong signatures of ram-pressure stripping also possess an active galactic nucleus (AGN), suggesting a possible correlation between the two phenomena. This result has not been confirmed by a subsequent study. 
Building upon previous findings, here  we combine MUSE observations conducted within the GASP program and a general survey of the literature to robustly measure the AGN fraction in cluster's ram pressure stripped galaxies using BPT emission line diagrams. Considering a sample of 115 ram pressure stripped  galaxies with stellar masses $\geq 10^9 \, M_{\odot}$, we find an AGN fraction of $\sim27\%$. This fraction strongly depends on stellar mass: it raises to 51\% when only  ram-pressure stripped galaxies of masses $M_* \geq 10^{10} \, M_{\odot}$ are considered. We then investigate whether the AGN incidence is in excess in ram pressure stripped galaxies compared to non-stripped galaxies, using as comparison a  sample of non-cluster galaxies observed by the survey MaNGA. Considering mass-matched samples, we find that the incidence of AGN activity is significantly higher (at a confidence level $>99.95\%$) when ram-pressure stripping is on act,
supporting the hypothesis of an AGN-ram pressure connection.

\end{abstract}



\section{Introduction} \label{sec:intro}

Both theoretical and observational studies concur that there is a strong connection between the presence of an Active Galactic Nucleus (AGN) and the host galaxy properties \citep[see][and references therein]{kormendy+2013}, suggesting that internal processes might regulate the AGN activity and, conversely, the AGN activity might be relevant for shaping galaxy properties. 
AGN are preferentially found in more massive galaxies \citep[$M_* > 10^{9} M_\odot$, see e.g.][]{juneau+2011, sabater+2013, lopes+2017, pimbblet+2013, kauffmann+2003, sanchez+2018, decarli+2007, rodriguez+2017} and the mass of the host galaxy is the main parameter driving the level of AGN activity \citep{magliocchetti+2020}. However, it is  still debated if other factors, such as  dense galaxy environment  or galaxy clusters \citep[e.g.][]{pimbblet+2013}, have an impact on the presence of AGN in galaxies.

Despite the vast literature on this topic \citep[e.g.][]{kauffmann+2004, best+2007, silverman+2009,vonderlinden+2010, hwang+2012, sabater+2013, martini+2013, ehlert+2014, silverman+2015, coldwell+2017,lopes+2017, marziani+2017, gordon+2018, magliocchetti+2018,koulouridis+2018, argudo-fernandez+2018}, different studies have  reached quite opposite results, most likely due to the different techniques adopted to identify AGNs, select the samples and characterise the environment. Using a spectroscopic sample, \cite{dressler+1985} first suggested that the fraction of AGN in clusters ($\sim$ 1\%) is significantly lower than in the field ($\sim$5\%). 
Similarly,  \cite{lopes+2017}, identifying AGN  in the SDSS using optical emission lines and \cite{baldwin+1981} (BPT) diagrams, found that AGN favor  environments typical of the field, low mass groups or cluster outskirts. Using the same dataset but a different cluster sample, \cite{vonderlinden+2010} showed instead that the AGN fraction does not change as a function of environment, nor of clustercentric distance \citep[see also][]{miller+2003}. Similar conclusions were obtained by  \cite{martini+2007}, \cite{lehmer+2007}, \cite{sivakoff+2008}, \cite{arnold+2009}, exploiting X-ray data.
Yet, the radio AGN fraction seems to be much higher in clusters than in the field \citep[][]{sabater+2013, best+2007}. 

Considering local density as a proxy for environment, \cite{kauffmann+2004} found that AGN host galaxies with strong [O III] emission are twice as frequent in low density regions than in high density regions \citep[see also][]{miller+2003,montero-dorta+2009}. In contrast, \cite{amiri+2019} did not find any effect of the galaxy density on nuclear activity. \cite{sabater+2013} found that (at fixed mass) the prevalence of optical AGN is a factor of 2-3 lower in the densest environments \citep[see also][]{man+2019}, but increases by a factor of $\sim$2 in the presence of strong one-on-one interactions. 
\cite{gilmour+2007} showed that X-ray selected AGN lie predominantly in  moderate dense regions.

The expected connection between AGN incidence and properties with the environment has roots in the fact that the characteristics of AGN are strongly linked to the conditions of the available gas, which in turn can be affected by the galaxy environment. 
So any environmental specific physical mechanism that has the potential to affect the galaxy gas can impact the AGN activity. 
For example, mergers -- which most frequently happen in the field -- have frequently been cited as a method to fuel AGN \citep[e.g.][]{sanders+1988} and a number of morphological studies claim an excess of post-merger systems in their AGN samples \citep[][]{bahcall+1997, canalizo+2001, urrutia+2008, letawe+2010, smirnova+2010}.

Another process able to affect the gas supply in galaxies is ram pressure stripping (RPS, \citealt{gunn1972}). This is a mechanisms happening most efficiently in clusters and massive groups \citep{hester2006} and it is due to the pressure exerted by the intracluster medium (ICM) on the galaxy interstellar medium (ISM).  This interaction can produce many visible effects on the galaxy, such as alter its less bound gas, giving rise to wakes of stripped material departing from the main galaxy body  \citep{vangorkom2004, kenney+2004, poggianti+2017a, fumagalli+2014} and inducing a quenching of star formation \citep{vollmer+2001, tonnesen+2007, vulcani+2020}. Prior to complete gas removal, it has been observed that ram pressure can also increase the star formation rate in galaxies \citep[][]{crowl-kenney+2006, merluzzi+2013, vulcani+2018} and simulations support this finding \citep[][]{kronberger+2008, kapferer+2009, tonnesen+2009, bekki+2014}, suggesting that the increased pressure initially helps compress the gas and triggers increased star formation. The same mechanism that initially promotes star formation can also fuel  the AGN during the RPS process: gas can be funnelled towards the galaxy centres,  due to  gravitational instabilities and the spiraling towards the center of clumps that lose angular momentum \citep[][]{schulz+2001, tonnesen+2009, ramos-martinez+2018}. The funneling of gas towards the galaxy center can also ignite the central super-massive black hole (SMBH). Theoretical models \citep{tonnesen+2009} have indeed demonstrated that gas inflows can fuel the central AGN in ram-pressure stripped galaxies, possibly due to the presence of magnetic fields \citep{ramos-martinez+2018}. The enhanced accretion onto the black hole can then produce heating and outflows due to AGN feedback \citep{ricarte+2020}.

Therefore RPS might be simultaneously responsible for an enhanced AGN activity and the appearance of tails of stripped material. This scenario has been first proposed by \citet[hereafter P17b]{poggianti+2017b}  to explain the very high incidence (6/7) of AGN detected in a sample of galaxies strongly affected by RPS,  also called jellyfish galaxies \citep[see also][]{maier+2021}. That analysis is based on Integral Field spectroscopic data coming from the GAs stripping Phenomena in galaxies \citep[GASP,][] {poggianti+2017a}. Subsequent GASP studies have led to the identification of AGN-driven outflows \citep{radovich+2019} and a compelling case for AGN feedback in action \citep{george+2019}.

The \citetalias{poggianti+2017b} analysis is based on a small sample and importantly is only composed of jellyfish galaxies with very striking tails, and all massive galaxies. Thus, it leaves open the possibility that the AGN activity could be only related to the (rather short) peak phase of stripping and/or only to the galaxy mass regardless of RPS.

A subsequent study did not find a high incidence of AGN: \cite{roman-oliveira+2019} \footnote{Note added at the proofs stage: see also Boselli et al. arxiv 2109.13614 which appeared unrefereed on astroph on September 28 2021.} analyzed a sample of ram pressure stripped galaxies
in a supercluster at $z\sim 0.2$ and found only 5/70 AGN, according to optical line diagnostics. Their sample  span a wide galaxy stellar mass range (from 10$^9$ to 10$^{11.5} M_\odot$) and is based on visual identification of the candidates. At odds with GASP, none of these candidates have IFS data to confirm they are indeed affected by RPS. It also includes ram pressure stripped candidates with different degrees of stripping, while as said above the \citetalias{poggianti+2017b} study includes only very dramatic cases.

In this paper, we build on previous results and aim at estimating the incidence of AGN on the largest possible sample of ram pressure stripped galaxies up to date. 

Throughout the paper, we adopt a \cite{chabrier+2003} initial mass function (IMF) in the mass range 0.1-100 M$_{\odot}$. The cosmological constants assumed are $\Omega_m=0.3$, $\Omega_{\Lambda}=0.7$ and H$_0=70$ km s$^{-1}$ Mpc$^{-1}$. 

\section{DATASETS and galaxy samples} \label{sec:highlight}

In this paper we aim at characterizing the incidence of AGN activity among ram pressure stripped cluster galaxies. We start our analysis  by considering a sample of galaxies drawn from the GAs Stripping Phenomena in galaxies (GASP, \citealt{poggianti+2017a}) survey, then, to increase the statistics and support more robustly the results, we also gather a literature sample of all ram pressure stripped galaxies identified by different authors in the last four decades. Finally, we exploit the fifteenth data release of the MaNGA survey (DR15; \citealt{aguado+2019}) to build a control sample of galaxies with characteristics similar to the ram pressure stripped galaxies, but considering only those {\sl not} in clusters and therefore presumably not affected by strong RPS.

\subsection{The GASP sample and data} \label{gasp sample}
GASP is a project aimed at studying gas removal processes, mainly due to ICM-ISM interaction, using a sample of 114 
galaxies. More specifically, it comprises both RPS candidates and undisturbed galaxies located in clusters, groups and field, spanning a range in stellar masses from $10^{9}$ to $3.2 \times 10^{11}$ M$_\odot$ and a redshift range of 0.04 $<$ z $<$ 0.07. 
All galaxies were selected on the basis of B-band imaging coming from  three different surveys: WINGS \citep{fasano+2006}, OMEGAWINGS \citep{gullieuszik+2015}, and PM2GC \citep{calvi+2011}. 

GASP is based on an ESO Large Program carried out with the integral-field spectrograph MUSE, mounted at the VLT, whose large field of view ($1'\times1'$) and high, but seeing limited, spatial resolution ($0.2^{\prime\prime}$/pixel, seeing of $1^{\prime\prime}$), allow us to cover the galaxy outskirts and possible tails of gas departing from the main body of the galaxies up to ten times the galaxy effective radius (i.e. $\sim \ 10 \ R_{e}$) with a resolution of $\sim$ 1 kpc at the galaxy redshifts.

For our analysis, we  select only cluster members that have been confirmed to be ram pressure stripped based on the MUSE data: in fact, they all have extraplanar $\rm H\alpha$ emission in various stages of stripping (B. Poggianti et al. in prep.), from weak/initial stripping (JStage=0.5) to significant tails (JStage=1) to extreme tails longer than the stellar disk diameter (JStage=2, so-called ``jellyfish galaxies'') to truncated disks corresponding to a late-stage of RPS (Jstage=3), for a total of 51 galaxies. From now on we will call this sample GASP-RPS. All of these are morphologically late-type and star-forming galaxies. 

We make use of the fluxes of the emission-only component of the lines H$\alpha$, H$\beta$, $\rm [OIII]\lambda$5007\AA \ and $\rm [NII]\lambda$6583\AA \ measured with the KUBEVIZ code \citep{fossati+2016} from the continuum-subtracted MUSE cubes corrected for both Galactic and intrinsic extinction, as described in detail in \cite{poggianti+2017a}.  Stellar masses are taken from \cite{vulcani+2018} and are computed using the SINOPSIS spectrophotometric code \citep{fritz+2017} by summing up the masses of all the spaxels within the galaxy disk \citep{gullieuszik+2020}.

To  characterize the ionization mechanism acting on the gas and therefore identify galaxies with AGN, we inspect the BPT diagnostic diagram [NII]/H$\alpha$ vs. [OIII]/H$\beta$ ratios (BPT-NII, \citealt{baldwin+1981}). We  consider only spaxels with a signal-to-noise (S/N) ratio greater than $3$ for all the lines used. We  use the relation from \cite{kauffmann+2003} (K03) to separate Star-forming from Composite regions,  the \cite{kewley+2001} (K01) line to identify AGN\footnote{The choice of the K01 demarcation line to identify AGN is a conservative choice that minimizes the AGN spaxels. Note that recent works (see \citealt{law+2021} and references therein) find a demarcation line in the NII-BPT diagram closer to the K03 separation, but we choose the most conservative one to minimize the contamination from star-forming regions.} and the \cite{sharp+2010} (SB10) relation to discriminate between Seyferts and LINERs. 
We classify a galaxy as AGN host if in its central (3$^{\prime\prime}$) region there are at least 20 spaxels\footnote{We adopted this number upon visual inspection of the maps. This choice allows us to have enough spaxels to identify possible AGN with high confidence, still focusing on the central part of the galaxy.} that have a Seyfert or LINER classification, otherwise we flag it as star forming. In the galaxies with high values of extinction ($A_V$, as measured by the Balmer decrement) in the central cores, that might prevent us from identifying an AGN, we further inspect the LINER classified spaxels:  a bi-conical shape of their distribution suggests extended ionized regions and therefore indicates the presence of the AGN. 

For those galaxies with a central LINER/AGN classification we have carefully checked the emission line fits, in particular the $\rm H\beta$ line given that an underestimate of H$\beta$ flux would lead to an overestimate of the $\rm [OIII]/H\beta$ ratio, mimicking line ratios typical of AGN.

\subsection{Ram-pressure stripping candidates from the literature}\label{sec:lit_sample}

We have performed a systematic literature search of all the ram pressure stripped galaxies identified  by December 2020. These galaxies were studied exploiting a wide variety of 
observational techniques, including radio \citep[e.g.,][]{gavazzi+1995}; sub-mm \citep[e.g.][] {Scott2013,jachym+2014, jachym+2019}; infrared \citep[e.g.,][]{sivandam+2010, sivanandam+2014}; optical \citep[e.g.,][]{gavazzi+1995,gavazzi+2001, sun+2007, sun+2010,yagi+2010,sivandam+2010, fumagalli+2014, gavazzi+2017, fossati+2016, roberts+2020}; UV (e.g., \citealt{smith+2010}) and X-ray \citep[e.g.,][]{sun+2006,sun+2010}.
The assembled sample is therefore greatly heterogeneous and while for some galaxies it has been confirmed that RPS is the only acting mechanism, in some other cases galaxies are most likely undergoing both RPS and tidal interactions. As our aim is to
include all RPS galaxies and collect a sample as large as possible, we consider also the latter cases.
However, we remove cases where a merger or tidal interaction is the main cause of the galaxy transformation.
\footnote{ The 
candidate merging systems removed from the sample are: F0237 \citep{owers+2012}, NGC4294, NGC4299 and NGC4302 \citep{vollmer+2013, pappalardo+2012}.}

We narrow down our search to galaxies for which we retrieve information at any wavelength on the ionization mechanism of the central emission,\footnote{A list of all RPS galaxies known in the literature, regardless of AGN information, will be published in J. Crossett et al. (in prep.).} obtaining a total sample of 80 galaxies (from now on LIT-RPS sample). All these turn out to have some active star formation (in addition to the eventual AGN activity) in the available literature.

The LIT-RPS sample is located in the redshift range $0.001 \leq z \leq 0.34$, plus a galaxy at $z=0.73$, and covers a stellar mass range of $1.3 \times 10^{8} < M_*/M_\odot < 2.0 \times 10^{11}$. Stellar masses have been collected from the literature and homogenized to the same \cite{chabrier+2003} IMF (as in GASP). When a stellar mass estimate was not available (4/80), we 
computed it using the available photometric data following  the \cite{bell_dejong+2001} approach, as described in Appendix \ref{sec:appendix_mass}. 

To assess the strength of RPS signatures and compare with the GASP JStage classification, four of us (GP, BMP, BV, AM) visually inspected the available images in the literature for all the galaxies. 
Following the scheme described in \S2.1, we assign a flag indicating the extent of the tail (JStage) based on the H$\alpha$ emission (if available) and also a general JStage
based on any wavelength observed (JStage$_{gen}$). In the case of multiple images with different resolutions or at different wavelengths showing a different extent of the tail, we always consider the wavelength with the longest visible tail to assign the JStage$_{gen}$. 
The classifiers agreed in most of the cases. In the discrepant cases, each galaxy was inspected together by the classifiers to ensure homogeneity and to find a consensus. This visual inspection also confirmed that the LIT-RPS sample is composed of morphologically late-type galaxies (spirals or irregulars).

\subsection{The MaNGA sample} \label{manga sample}

MaNGA (Mapping Nearby Galaxies at Apache Point Observatory, \citealt{bundy+2015}) is an integral-field spectroscopic survey observing galaxies at 0.01 $\leq$ z $\leq$ 0.15 using the BOSS Spectrograph \citep{smee+2013} mounted at the 2.5 m SDSS telescope \citep{gunn+2006}, which covers a spectral range from 3600 \AA \ to 10300 \AA, with a resolution of R$\sim$ 1400 at 4000 \AA \ and R$\sim$ 2600 at 9000 \AA. 

We exploit the MaNGA DR15 release and use the outputs of the Pipe3D pipeline \citep{sanchez+2016,sanchez+2016b, sanchez+2018}. More specifically, we use the Pipe3D-v2\_4\_3\footnote{\url{https://www.sdss.org/dr16/data_access/value-added-catalogs/?vac_id=manga-pipe3d-value-added-catalog:-spatially-resolved-and-integrated-properties-of-galaxies-for-dr15}} catalog, which contains integrated properties, characteristic and gradients of different quantities for 4656 galaxies. Of interest for our work are integrated stellar masses and star formation rates obtained from the H$\alpha$ emission line, that we convert to our adopted \cite{chabrier+2003} IMF.

We first exclude from the sample 75 duplicate galaxies and then select only galaxies with a specific Star Formation Rate (sSFR) $> 10^{-11}\rm yr^{-1}$, for a total of 2509 galaxies. The latter selection allows us to consider only star forming galaxies, as are galaxies in both the  GASP-RPS and LIT-RPS samples.

As we aim at assembling a sample not affected by RPS, we crossmatch our sample with the environmental catalog by \cite{tempel+2014}, who provide halo mass estimates based on \cite{navarro+1997} profiles. Using a searching radius of 5$^{\prime\prime}$, we obtain a match for 2061 galaxies, 
861 of which are located in structures with halo masses $\log(\rm M_h/M_\odot) < 13.0$,\footnote{We verified that results are insensitive to the exact choice of this threshold, exploring log halo masses up to 13.6. We decided to use a conservative cut (13.0) to avoid the possibility that ram-pressure stripped galaxies in groups contaminate the sample.} therefore are most likely isolated \citep{yang+2007}.

Finally, to reduce the effect of a different morphological mix among the different samples, we use the visual morphological classification from the MaNGA Value Added Catalogs\footnote{\url{https://www.sdss.org/dr16/data_access/value-added-catalogs/?vac_id=manga-visual-morphologies-from-sdss-and-desi-images}} that is based on inspection of image mosaics using a new re-processing of SDSS and Dark Energy Legacy Survey (DESI) images, following the methods from \cite{hernandez+2010} and exclude 
70 early-type (Ellipticals, S0s and S0as) and 
2 unclassified galaxies.\footnote{For consistency with the other samples, we have applied the morphological cut, but all the results remain unchanged if no morphological criterion is applied.} 

To assemble the final MaNGA sample, we consider only galaxies that in a circular aperture of  3$^{\prime\prime}$ diameter (i.e. the SDSS fibre size) centered on the galaxy have at least 20 spaxels with $\rm S/N > 3$ for all lines that will be used to detect the presence of an AGN. 
782 galaxies pass this selection and constitute our reference sample, called MaNGA-Ref. This sample covers a redshift range 0.0024$<z<$0.1439.
 We note that the MaNGA fibre core diameter (2$^{\prime\prime}$) is similar to the typical seeing value (2$^{\prime\prime}$.5). At the median redshift of our MaNGA-Ref sample (z=0.0317), the MaNGA spatial resolution of 2$^{\prime\prime}$ corresponds to 1.27 kpc and at the 75\% redshift percentile (z=0.043) to 1.7 kpc. 
This is only slightly worse than the GASP spatial resolution of $\sim$1 kpc  (median $z = 0.05$), which is dominated by the seeing ($\sim$1”).\footnote{ This is why in GASP the datacubes have been filtered with a 5x5 kernel \citep{poggianti+2017a}.}


To identify AGN, we inspect the  BPT-NII maps provided by the online tool MARVIN\footnote{\url{https://www.sdss.org/dr15/manga/marvin/}} and use the same classification criteria as for the RPS samples.
We count the number of spaxels classified as AGN (i.e. Seyfert + LINER), Star-Forming or Composite.
If the number of spaxels classified as Seyfert or LINER is larger than 20 in a circular aperture of 3$^{\prime\prime}$ diameter, we classify the galaxy as AGN, otherwise as star forming. 

\section{Results I: The incidence of AGN among ram pressure stripped galaxies}
In this section we present the sample of AGN hosts  in the GASP-RPS and in the LIT-RPS samples separately and quantify the incidence of AGN among ram pressure stripped galaxies. We also investigate if these fractions depend on the properties of the ram pressure stripped galaxies, such as stellar mass and Jstage. 
In the following section we will quantitatively compare these fractions controlling for the different mass distribution and comparing them to those of the MaNGA-Ref sample.

\subsection{GASP-RPS} \label{sec:GASP_res}

\begin{figure*}[!t]
\centering
\includegraphics[scale=0.5, trim={3cm 0 5cm 0}]{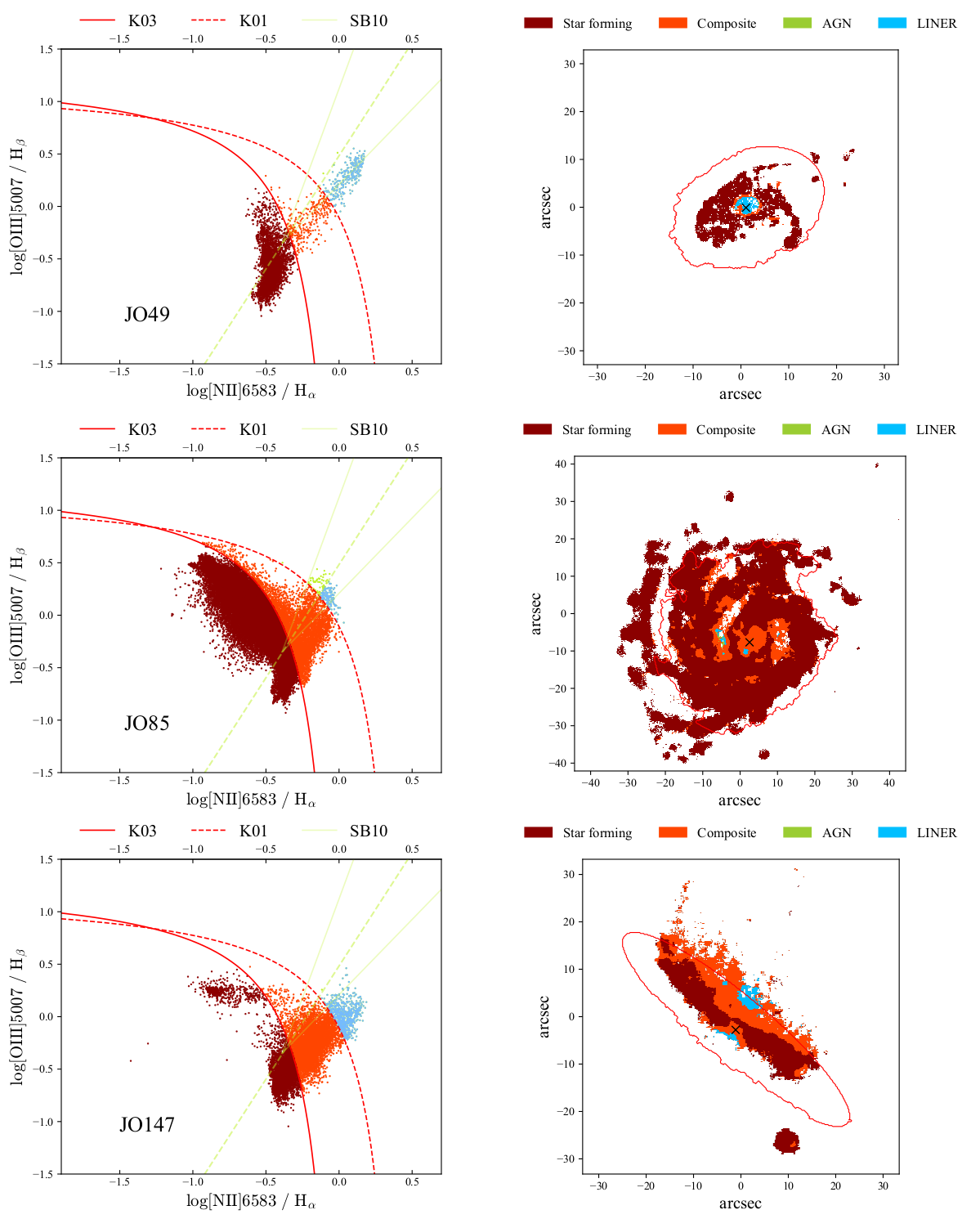}
\caption{Left. BPT-NII diagnostic diagram for all spaxels with $S/N > 3$. The red dotted and continuous lines are defined as in \citet{kewley+2001} and \citet{kauffmann+2003},  respectively. The green lines are taken from \citet{sharp+2010}. Right. Galaxy map color-coded according to the BPT-NII classification; red lines are the stellar emission isocontours corresponding to the galactic disk edges. \label{bpt1}}
\end{figure*}

\begin{figure*}[!t]
\figurenum{1} 
\centering
\includegraphics[scale=0.5, trim={0cm 2cm 5cm 0}]{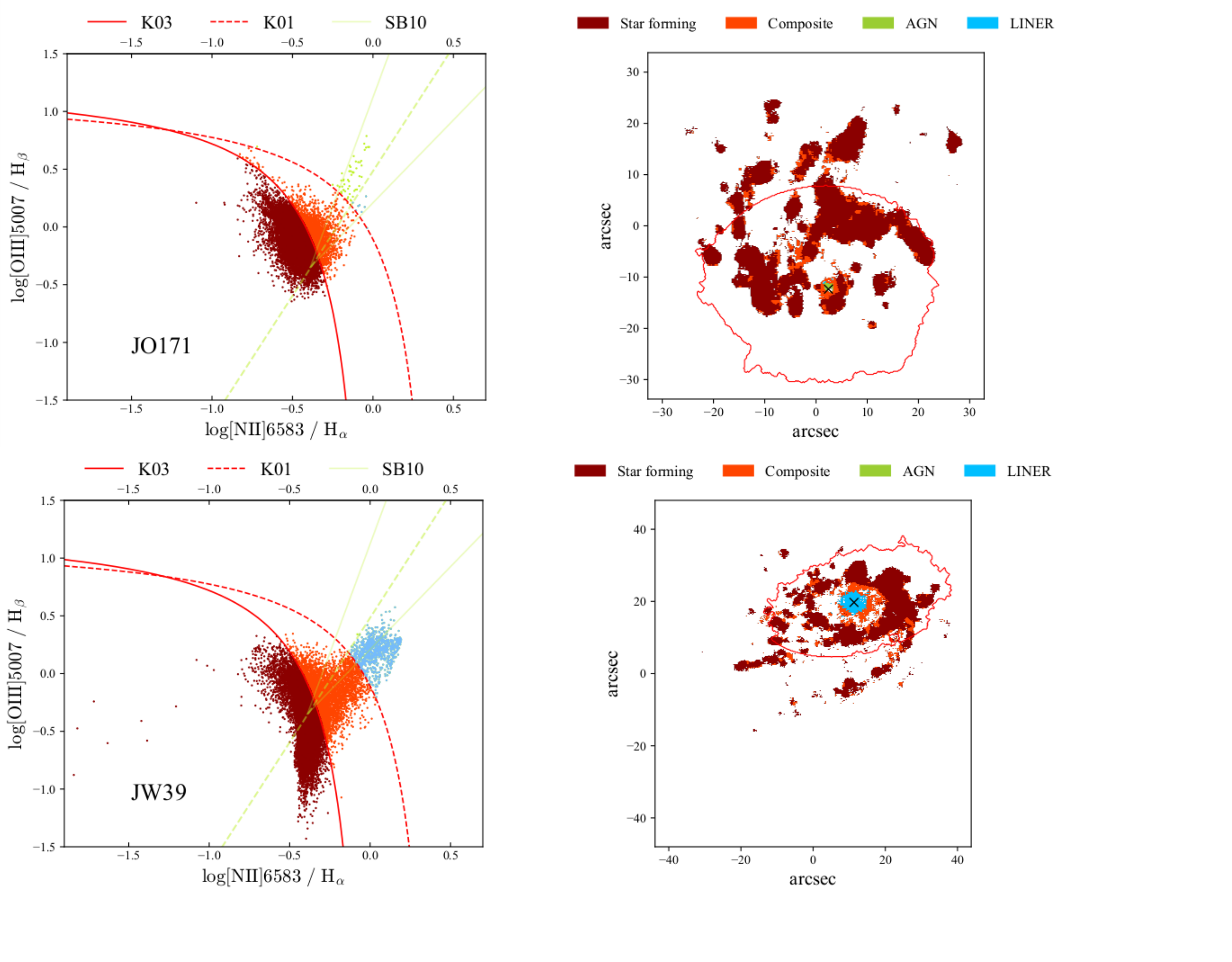}
\caption{{\it (continued)}}
\end{figure*}

In the GASP-RPS sample, seven galaxies are already known to host an AGN; six of them were presented in \citetalias{poggianti+2017b}  and one (JO36), in \cite{fritz+2017}. The latter is an edge-on disk hosting an obscured AGN which is not directly identified using BPT diagrams due to strong dust absorption. However, evidence for the AGN presence comes from extra-nuclear LINER-like emission with a cone morphology and the AGN is detected by \emph{Chandra} as a point-like X-ray source \citep{fritz+2017}.

Among the \citetalias{poggianti+2017b} candidates, JO194 was classified as a LINER and its combined line ratios are better reproduced by an AGN model \citep{radovich+2019}, while JO201, JO206, JO204, JW100 and JO135 are classified as Seyfert galaxies according to BPT diagrams and are all Seyfert2. JO204 and JO135 also have extended emission line regions ionized by the AGN. Four of these galaxies display AGN outflows \citep{radovich+2019}.

Having inspected all other GASP cluster-members ram-pressure stripped galaxies, we find other 5 AGN candidates that are presented  in Figure \ref{bpt1}. The stripping characteristics of these galaxies were discussed in previous works and only summarized here, but the analysis of their central ionization mechanism is shown here for the first time. 
JO49 has unwinding tails due to RPS \citep{bellhouse+2021}. It presents a central LINER-like region surrounded by a thin Composite-like ring, which in the BPT diagram correspond to a long finger of points encompassing the Composite region extending well beyond the K01's line. We note that JO49 hosts also an X-ray source of luminosity $1.2 \times 10^{41} \rm \, erg \, s^{-1}$, detected by XMM-Newton (\citealt{webb+2020}\footnote{The 4XMM-DR10 catalog contains source detections covering an energy interval from 0.2 keV to 12 keV. On the other hand, the \emph{Chandra} energy range goes from 0.5 to 7 keV.}).

JO85, another unwinding ram pressure stripped galaxy \citep{bellhouse+2021},  has fewer LINER-like points than JO49 embedded in a Composite-like region, but it is highly obscured by dust ($A_V \sim 2.7$ in the central region as measured by the Balmer decrement) and has a central \emph{Chandra} point source with a luminosity of $5.0 \times 10^{40} \, \rm erg \, s^{-1}$ \citep{evans+2020}.

JO147 (first described by \citealt{merluzzi+2013}, see also \citealt{poggianti+2019}) is an inclined highly-extincted disk, and is stripped in the north-west direction. We find that it has LINER-like opposite
cones embedded in wider Composite cones. Its luminosity in the X-ray band observed by XMM-Newton is
$2.4 \times 10^{41} \rm \, erg \, s^{-1}$ \citep{webb+2020}.

JO171 is an Hoag-like ring galaxy with long tails stripped in the north direction \citep{moretti+2018}. 
It has central AGN-powered spaxels (Seyfert2) in the inner kpc. 

Finally, JW39  has long tails originating from unwinding spiral arms \citep{bellhouse+2021}. It has a LINER-like circular central region surrounded by a 
larger circular area with Composite emission.

The latter two galaxies have no available central X-ray counterparts,  from neither XMM-Newton nor \emph{Chandra}.

To summarize, with respect to the sample of AGN described in \citetalias{poggianti+2017b} and \cite{radovich+2019}, we find an additional Seyfert2 and 4 LINER-like galaxies, yielding a total sample of 12 AGN hosts in the GASP-RPS sample.
Their main properties 
are summarized in Tab. \ref{gasp_agn}. 

\begin{figure*}
    \makebox[\textwidth]{
    \centering
    \includegraphics[scale=0.55]{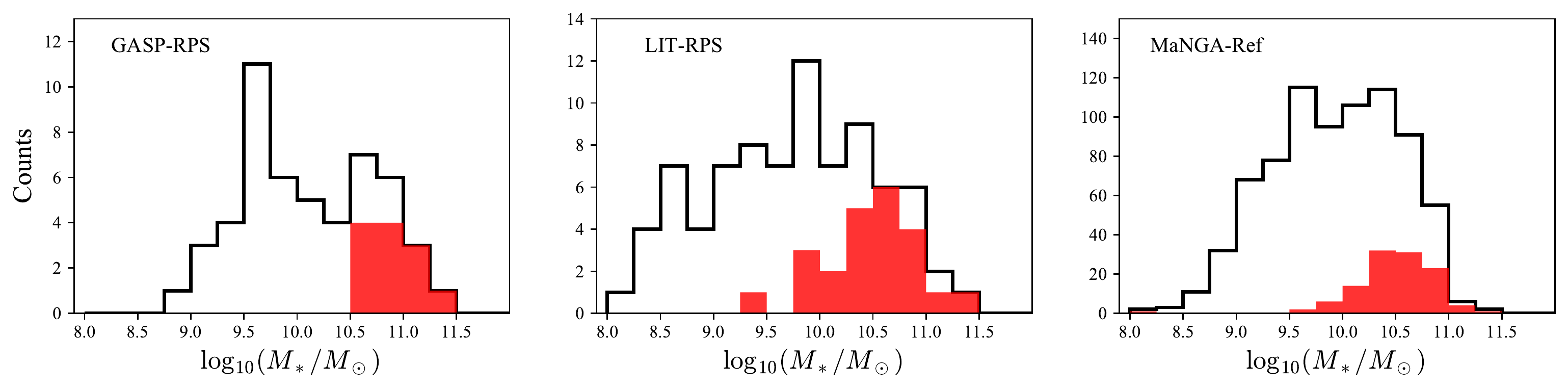}}%
    \caption{Stellar mass distributions for all galaxies (black histogram) and for galaxies hosting an AGN (red histogram). From left to right: the GASP-RPS, LIT-RPS and MaNGA-Ref samples. \label{massdistr}}
\end{figure*}

The left panel of Figure \ref{massdistr} shows the mass distribution of galaxies hosting an AGN compared to the entire GASP-RPS sample. While RPS galaxies cover a mass range of 8.7 $ \leq \log(M_\ast/M_\odot) \leq$ 11.5, AGN hosts are among the most massive galaxies in the sample, having all $\log(M_\ast/M_\odot) \geq$ 10.5.

We are now in the position of computing the fraction of AGN ($f_{AGN}$) over the total (AGN+SF) number of galaxies, considering different subsamples, as summarised in   Table \ref{fagn}.
The AGN fraction in the total GASP-RPS sample is 0.24$^{+0.06}_{-0.05}$, with uncertainties computed as binonomial errors.
Restricting the sample to $\log(M_\ast/M_\odot) \geq$ 10.5,
this fraction becomes 0.71$^{+0.10}_{-0.12}$.

\begin{figure*}
    \makebox[\textwidth]{
    \centering
    \includegraphics[scale=0.7]{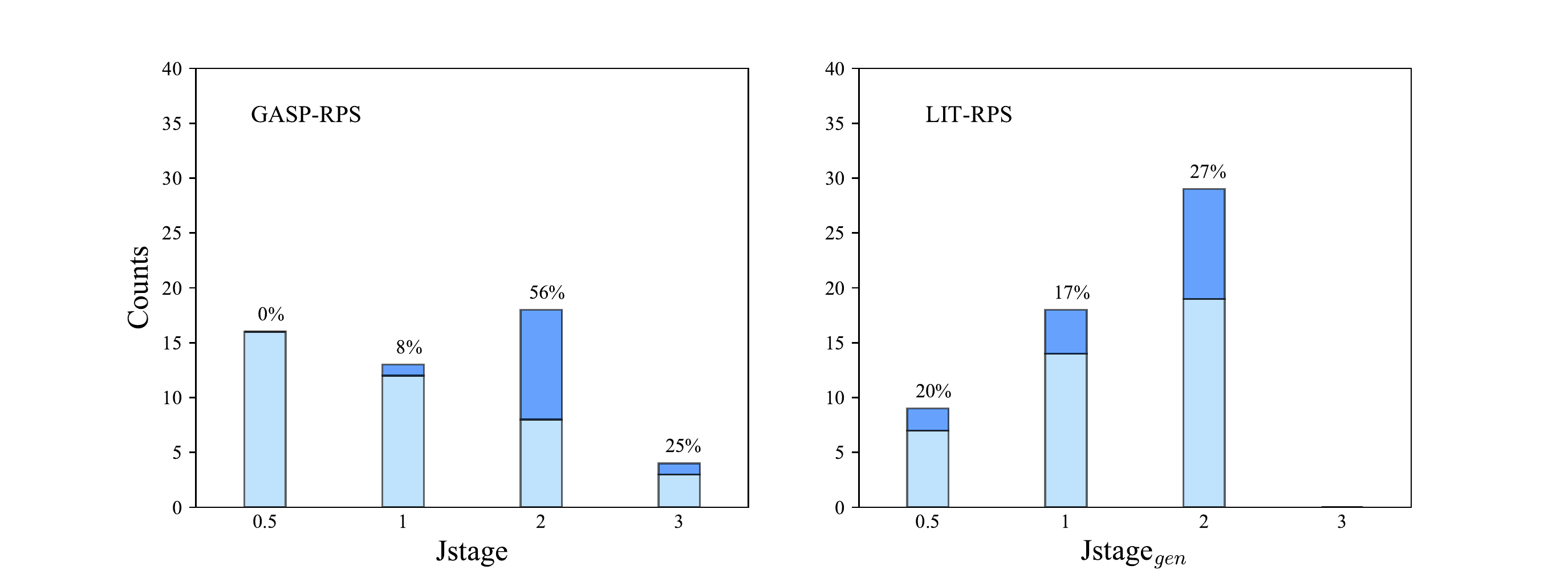}}%
    \caption{Left. Stacked Histograms for galaxies of different Jstages and divided among centrally star-forming galaxies (light blue histogram) and AGN (dark blue histogram) according to the BPT-NII classification for GASP-RPS (left) and LIT-RPS (right). Percentages are AGN fractions in the corresponding bin of Jstage and Jstage$_{gen}$.\label{hist_jphase}}
\end{figure*}

Considering the various stages of stripping (Fig. \ref{hist_jphase}), the frequency of AGN increases with the strength of RPS signatures:
no galaxies with AGN activity have Jstage $= 0.5$, while the AGN incidence increases among moderate-stripping galaxies (Jstage=1,  8\%) and 
is particularly high among  Jstage=2 galaxies, where it reaches 56\%.
Still one out of 4 galaxies in a late stage of RPS (truncated disks, Jstage=3) has an AGN.
Interestingly, using the same methods of the current analysis, only two AGNs are found in the GASP non-ram pressure stripped sample of star-forming galaxies, which consists of 49 galaxies \citep[][B. Poggianti et al. in prep.]{vulcani+2021}.

In the GASP-RPS sample, out of the 17 galaxies more massive than $\log(M_{*}/M_\odot) \geq 10.5$, 10 ($\sim 58 \%$) have 
Jstage=2 and, viceversa, $\sim 55 \%$ (10/18) of the Jstage=2 galaxies are more massive than $\log(M_{*}/M_\odot) \geq 10.5$. All of them host an AGN. It is significant that none of the massive galaxies have a Jstage=0.5.
This result suggests a tight correlation between stellar mass and Jstage. 
The correlation is probably linked with the higher capability of massive galaxies to retain gas.
While low-mass galaxies are already completely stripped when they approach the densest regions in clusters, high-mass galaxies more easily hold onto their gas \citep[][Luber et al., submitted to ApJ]{jaffe+2018} and 
experience RPS in these dense regions, where the gas removal is the most intense.
Since AGN are preferentially located in the most stripped and massive galaxies, 
we cannot state which of these two parameters is more connected to the presence of an AGN.

\begin{deluxetable*}{ccccccccc}
\tablecaption{AGN candidates in the GASP sample. Columns are: 1) GASP ID; 2-3) coordinates of the optical center; 4) galaxy redshift; 5) host cluster; 6) galaxy stellar masses \citep{vulcani+2018}; 7) Jstage (Poggianti et al. in prep.; 8) AGN classification; 9) 
work in which the source is presented. 
\label{gasp_agn}}

\tablehead{
  \colhead{ID} &
  \colhead{RA} &
  \colhead{DEC} &
  \colhead{z} &
  \colhead{cluster} &
  \colhead{$\log M_\ast/M_\odot$} &
  \colhead{Jstage} &
  \colhead{AGN flag} &
  \colhead{refs}
 }
\startdata
\hline
JO85 & 351.13068 & 16.86815 & 0.0355 & A2589 & 10.7 & 1 & 3 & this paper\\
JO36 & 18.247583 & 15.591488 & 0.0407 & A160 & 10.8 & 3 & 4 & Fritz et al. (2017)\\
JO194 & 359.25284 & -34.680588 & 0.042 & A4059 & 11.2 & 2 & 3 & \citetalias{poggianti+2017b}\\
JO204 & 153.44513 & -0.914182 & 0.0424 & A957 & 10.6 & 2 & 1 & \citetalias{poggianti+2017b}\\
JO201 & 10.376208 & -9.26275 & 0.0446 & A85 & 10.8 & 2 & 1 & \citetalias{poggianti+2017b}\\
JO49 & 18.682709 & 0.286136 & 0.0451 & A168 & 10.7 & 2 & 3 & this paper\\
JO147 & 201.70721 & -31.395975 & 0.0506 & A3558 & 11.0 & 2 & 3 & this paper\\
JO206 & 318.44754 & 2.476218 & 0.0511 & IIZW108 & 11.0 & 2 & 1 & \citetalias{poggianti+2017b}\\
JO171 & 302.56125 & -56.641823 & 0.0521 & A3667 & 10.6 & 2 & 1 & this paper\\
JO135 & 194.26791 & -30.375088 & 0.0544 & A3532 & 11.0 & 2 & 3 & \citetalias{poggianti+2017b}\\
JW100 & 354.10443 & 21.150702 & 0.0619 & A2626 & 11.4 & 2 & 3 & \citetalias{poggianti+2017b}\\
JW39 & 196.03212 & 19.210691 & 0.0663 & A1668 & 11.2 & 2 & 3 & this paper\\
\hline
\enddata
\tablecomments{The adopted AGN flag for both GASP-RPS and LIT-RPS galaxies ranges from 0 to 6: 0 means that star formation is the dominant ionization process at the galaxy center according to BPT-NII classification; 1, 2, 3  if the galaxy hosts a Seyfert 1, Seyfert 2 or LINER-like nucleus, respectively, again according to the BPT diagram; 4  if the AGN has been detected through the X-ray signal, but not in the optical; 5  when the galaxy is classified as a radio galaxy; 6  when the source is classified as AGN, without any specification on the type.}
\end{deluxetable*}

\begin{table}
\caption{AGN fractions in the GASP-RPS sample, considering galaxies of different mass and characterized by different Jstages. Errors are binomial. 
\label{fagn}}
\scriptsize
\begin{tabular}{cccc}
\hline\hline
{$N_{AGN}/N_{TOT}$} &
{$f_{AGN}$} &
Jstage &
{$\log (M_\ast/M_\odot)$} \\
    \hline
    12/51 & $0.24_{-0.05}^{+0.06}$ & $\geq$ 0.5 & all \\ 
    12/17 & $0.71_{-0.12}^{+0.10}$ & $\geq$ 0.5 & $\geq$ 10.5 \\
    0/16 &  $0.0_{-0.0}^{+0.06}$ & $=0.5$ & all\\
    1/13 & $0.08_{-0.05}^{+0.11}$ & $=1$ & all\\
    10/18 & $0.56_{-0.12}^{+0.11}$ & $=2$ & all\\
    1/4 &  $0.25_{-0.15}^{+0.25}$ & =$3$ & all\\
    \hline
\end{tabular}
\end{table}

\subsection{LIT-RPS}

The catalog of the 82 literature ram-pressure stripped galaxies with SF/AGN information is presented in Table \ref{litt_sample}, 
which gives galaxy name, coordinates, redshift, host cluster name, Jstage, Jstage$_{gen}$ and alternative names. Stellar masses and AGN classification, 
along with the source for those values are given in Table \ref{litt_sample_ref}.

One of these galaxies has broad optical lines typical of Seyfert1.
For all the other galaxies the AGN classification is based on the BPT-NII diagnostic. 
For $\sim 68\%$ (55/81) of them the classification in published results is based on spectroscopic observations published in dedicated papers,  either 
from integral-field unit \citep{merluzzi+2013, merluzzi+2016, fossati+2016, boselli+2019, stroe+2020, consolandi+2017}, long-slit or fibre spectra (\citealt{ebeling+2019, cortese+2007, owers+2012, owen+2006,mahajan+2010}\footnote{We note that for 2 galaxies, GMP3618 and D100, \cite{mahajan+2010} give different results with respect to the classification reported in the DR7 \citep{abazajian+2009} and DR8 \citep{aihara+2011} analysis 
even though they used DR7 data to build up BPT-NII.}, \citealt{veron+2003}). 
For the other 32$\%$ (26/81) of the 
galaxies we instead use the online AGN classification based on the analysis of emission line ratios extracted from an integrated spectra of the central circular aperture ($r \sim 3^{\prime\prime}$) observed with the SDSS fibre (Data Release 8, from now on DR8, \citealt{aihara+2011}) as analyzed by  \cite{brinchmann+2004}, \cite{kauffmann+2003}, and \cite{tremonti+2004} in the Value Added Catalog MPA/JHU. 
For 24 galaxies, we had both the DR8 automatic classification and information about the central source from individual publications in the literature. In these cases we favoured the latter. 

We note that 4 of the 81 galaxies also have information coming from  either X-ray or radio data \citep{ebeling+2019, winkler+1992, owers+2012, best+2012, kalita+2019, caglar+2020}. While their position on a BPT-NII diagram suggests they are star forming, the additional data instead classify them as AGN. In what follows we will therefore discuss how results change if we include or exclude these 4 objects.

Overall, 24/82 galaxies host an AGN ($\sim $30\%). If we disregard the AGN classification based on  X-ray or radio data and consistently consider only the BPT-NII classification, the fraction above becomes 20/82 ($\sim 24 \%$).

The central panel in  Fig.\ref{massdistr} shows the mass distribution of the galaxies with and without AGN. The entire sample spans a mass range 8.1 $< \log(M_\ast/M_\odot)<$11.4. Similarly to what found for GASP, most of the AGN are massive galaxies, even though in this sample there are also a few less massive AGN hosts.
Above the GASP AGN mass limit ($\log(M_\ast/M_\odot) > 10.5$) the AGN fraction becomes 0.80$^{+0.08}_{-0.12}$.

Table  \ref{fagn_litt} reports the AGN fraction for the different subsamples considered, including that for galaxies of different Jstage$_{gen}$. The trend of the AGN fraction with Jstage$_{gen}$  is weaker than in GASP-RPS (see also Fig. \ref{hist_jphase}), with the percentages ranging between 17\% and 27\% but being consistent within the large errors in all Jstages$_{gen}$.\footnote{The subsample of LIT-RPS galaxies with Jstage is too small to study trends with the length of the H$\alpha$ tails.} 

We remind the reader that while the AGN classification and mass estimates among GASP galaxies are homogeneous, for the LIT-RPS sample we based the former on a number of different data and  indicators.
In addition, stellar masses have been computed following many different approaches and so, even though homogenized to the same IMF, there could be some systematics among the different galaxies. Finally, we recall that the Jstage$_{gen}$ flag is based on a very heterogeneous set of images in terms of wavelengths, depth, quality and therefore  results must be taken with caution.

\begin{table*}
\begin{center}
\caption{Colums are: 1) galaxy most common name 2) and 3) equatorial coordinates of the galaxy center from SIMBAD 4) galaxy redshift 5) host cluster 6) and 7) Jstage and general Jstage, defined in section \S3.2 8) alternative names. 
Stellar masses flagged with the asterisk (*) are computed by means of photometric data as described in Sec.\ref{sec:appendix_mass}. \label{litt_sample}} 
\scriptsize
\begin{tabular}{cccccccc}
\hline
\hline
name & RA & DEC & z & cluster & Jstage   &Jstage$_{\rm gen}$ & alternative names \\
\hline
  MIP001417-302303 & 3.5693 & -30.3843 & 0.2955 & A2744 & -- & 2.0 & F1228\\
  HLS001427-302344 & 3.61065 & -30.39581 & 0.3033 & A2744 & -- & 2.0 & F0083\\
  NGC1566 & 65.00175 & -54.93781 & 0.005 & Dorado & 0.0 & 0.0 & --\\
  ID345 & 149.9191 & 2.5281 & 0.727 & CGr32 & 2.0 & 2.0 & --\\
  LEDA36382 & 175.73523 & 19.96621 & 0.02427 & A1367 & 2.0 & 2.0 & CGCG97073\\
\hline
\end{tabular}%
\end{center}
\tablecomments{This table is published in its entirety in machine-readable format.  A portion is shown here for guidance regarding its form and content.}\end{table*}

\begin{table*}
\begin{center}
\caption{Columns are 1) galaxy name 2) logarithm of the stellar masses, adopting  Chabrier 2003 IMF (in parenthesis the reference from which the value has been taken is reported) 3) AGN classification (see Table \ref{gasp_agn}) and relative reference 4) references which present a characterization of the galaxy as a RPS candidate. The four galaxies with classification equals to 4 and 5 (e.g. where the AGN is spotted observing them in X and radio) resulted to be Star-Forming in the optics. In the text we analyze the consequence to change their AGN flag to 0. 
\label{litt_sample_ref}} 
\scriptsize
\begin{tabular}{cccc}
\hline
\hline
name & $\log(M_\ast/M_\odot)$ & AGN & Refs  \\
\hline
  MIP001417-302303 & 9.6 [\cite{rawle+2014}] & 0 [\cite{owers+2012}] & \cite{owers+2012}, \cite{rawle+2014}\\
  HLS001427-302344 & 10.9 [\cite{rawle+2014})] & 1 [\cite{owers+2012})] & \cite{owers+2012}, \cite{rawle+2014}\\ 
  NGC1566 & 10.8 [\cite{elagali+2019}] & 1 [\cite{veron+2006}] & \cite{elagali+2019}\\
  ID345 & 10.3 [\cite{boselli+2019}] & 1 [\cite{boselli+2019}] & \cite{boselli+2019}\\
  \multirow{2}{*}{LEDA36382} & \multirow{2}{*}{9.5 [\cite{mendel+2014}]} & \multirow{2}{*}{0 [\cite{sdss}]} & \cite{gavazzi+1995}, \cite{gavazzi+2001},\\
  &&& \cite{sivanandam+2014}, \cite{boselli+2018}, \cite{yagi+2017} \\
\hline
\end{tabular}%
\end{center}
\tablecomments{This table is published in its entirety in machine-readable format.  A portion is shown here for guidance regarding its form and content.}\end{table*}

\let\clearpage\relax

\newcolumntype{P}[1]{>{\centering\arraybackslash}p{#1}}
\startlongtable
\begin{deluxetable*}{lP{4cm}P{4cm}P{4cm}}
\tablecaption{Columns are 1) galaxy most common name 2) logarithm of the stellar masses, homogenizing them to our adopted Chabrier IMF (in parenthesis the reference from which the value has been taken is reported) 3) AGN classification (see Table \ref{gasp_agn}) and relative reference 4) references which present a characterization of the galaxy as a RPS candidate. The four galaxies with classification equals to 4 and 5 (e.g. where the AGN is spotted observing them in X and radio) resulted to be Star-Forming in the optics. In the text we analyze the consequence to change their AGN flag to 0. 
\label{litt_sample_ref}}
\tablehead{
\colhead{Name} &
\colhead{$\log(M_\ast/M_\odot)$} &
\colhead{AGN} &
\colhead{Refs} 
}
\def\arraystretch{2}
\startdata
\hline
  MIP001417-302303 & 9.6 [\cite{rawle+2014}] & 0 [\cite{owers+2012}] & \cite{owers+2012}, \cite{rawle+2014}\\
  HLS001427-302344 & 10.9 [\cite{rawle+2014})] & 1 [\cite{owers+2012})] & \cite{owers+2012}, \cite{rawle+2014}\\ 
  NGC1566 & 10.8 [\cite{elagali+2019}] & 1 [\cite{veron+2006}] & \cite{elagali+2019}\\
  ID345 & 10.3 [\cite{boselli+2019}] & 1 [\cite{boselli+2019}] & \cite{boselli+2019}\\
  LEDA36382 & 9.5 [\cite{mendel+2014}] & 0 [\cite{sdss}] & \cite{gavazzi+1995}, \cite{gavazzi+2001}, \cite{sivanandam+2014}, \cite{boselli+2018}, \cite{yagi+2017} \\
  UGC6697 & 9.8 [\cite{yagi+2017}] & 0 [\cite{consolandi+2017}] & \cite{gavazzi+1984}, \cite{gavazzi+1989}, \cite{gavazzi+1995},\cite{gavazzi+2001}, \cite{scott+2010}, \cite{fruscione+1990}, \cite{sun+2005}, \cite{yagi+2017}, \cite{consolandi+2017}\\
  2MASXJ11443212+2006238 & 9.9  [\cite{yagi+2017}] & 0  [\cite{sdss}] & \cite{gavazzi+1989}, \cite{gavazzi+1995}, \cite{gavazzi+2001}, \cite{yagi+2017}, \cite{gavazzi+2017}\\
  NGC4254 & 10.1 [\cite{boselli+2018}] & 2 [\cite{davies+2020}] & \cite{boselli+2018}, \cite{wezgowiec+2012}, \cite{chyzy+2007}, \cite{kantharia+2008},  \cite{rahman+2011}\\
  NGC4388 & 9.9 [\cite{boselli+2015}] & 2 [\cite{boselli+2015}] & \cite{vollmer+2003}, \cite{vollmer+2009}, \cite{oosterlo+2005}, \cite{chung+2009}, \cite{yoshida+2004}, \cite{yoshida+2002}, \cite{yagi+2013}, \cite{gu+2013},\cite{damas+2016}\\
  NGC4402 & 10.0 [\cite{boselli+2015}] & 0 [\cite{decarli+2007}] &  \cite{lee+2017}, \cite{abramson+2014}, \cite{cramer+2020}, \cite{abramson+2016}\\
  NGC4424 & 10.2 [\cite{boselli+2015}]  & 0 [\cite{filippenko+1997}] & \cite{chung+2007}\\
  NGC4438 & 10.4 [\cite{boselli+2015}] & 3 [\cite{veron+2003}] & \cite{boselli+2005}, \cite{vollmer+2009}, \cite{wang+2020}, \cite{kenney+1995}, \cite{chemin+2005}, \cite{kenney+2008}\\
  IC3418 & 8.4 [\cite{boselli+2015}] & 0 [\cite{fumagalli+2011}] & \cite{hester+2010}, \cite{fumagalli+2011}, \cite{kenney+2014}\\
  NGC4501 & 10.7 [\cite{boselli+2015}] & 2 [\cite{boselli+2015}] & --\\
  NGC4522 & 9.1 [\cite{boselli+2015}] & 0 [\cite{sdss}] & \cite{kenney+2004}, \cite{abramson+2016}, \cite{lee+2018}, \cite{minchin+2019}, \cite{stein+2017}, \cite{vollmer+2004}\\
  NGC4548 & 10.5 [\cite{boselli+2015}] & 3 [\cite{boselli+2015}] & --\\
  NGC4569 & 10.4 [\cite{boselli+2015}] & 3  [\cite{boselli+2015}] & \cite{boselli+2016}, \cite{boselli+2006}, \cite{tschoke+2001}, \cite{vollmer+2004}, \cite{wezgowiec+2012}\\
  GMP6364 & 8.5 [\cite{salim+2016}] & 0 [\cite{sdss}] & \cite{roberts+2020}, \cite{salim+2018}, \cite{salim+2016}\\
  GMP5821 & 8.9 [\cite{salim+2016}] & 0 [\cite{sdss}] & \cite{roberts+2020}, \cite{salim+2018}, \cite{salim+2016}\\
  IC3913 & 11.0 [\cite{salim+2016}] & 0 [\cite{mahajan+2010}] & \cite{salim+2018}, \cite{salim+2018}, \cite{salim+2016}\\
  GMP5382 & 9.3 [\cite{salim+2016}] & 0 [\cite{sdss}] & \cite{roberts+2020}, \cite{salim+2018}, \cite{salim+2016}\\
  GMP4688 & 8.5 [\cite{salim+2016}] & 0 [\cite{sdss}] & \cite{roberts+2020}, \cite{salim+2018}, \cite{salim+2016}\\
  GMP4629 & 8.6 [\cite{salim+2016}] & 0 [\cite{sdss}] & \cite{salim+2018}, \cite{chen+2020}, \cite{roberts+2020}, \cite{gavazzi+2018}, \cite{salim+2018}, \cite{salim+2016}\\
  GMP4570 & 8.1 [\cite{salim+2016}] & 0 [\cite{mahajan+2010}] & \cite{salim+2018}, \cite{chen+2020}, \cite{roberts+2020}, \cite{gavazzi+2018}, \cite{salim+2018}, \cite{salim+2016}\\
  GMP4555 & 9.9 [\cite{salim+2016}] & 0 [\cite{mahajan+2010}] & \cite{salim+2018}, \cite{chen+2020}, \cite{salim+2018}, \cite{salim+2016}, \cite{gavazzi+2018}\\
  MACSJ1258-JFG1 & 10.5 (*) & 1 [\cite{rakshit+2017}] & \cite{ebeling+2014}, \cite{mcpartland+2016}\\
  NGC4848 & 10.8 [\cite{salim+2016}] & 0 [\cite{mahajan+2010}] & \cite{chen+2020}, \cite{salim+2018}, \cite{roberts+2020}, \cite{salim+2018}, \cite{salim+2016}, \cite{fossati+2012}, \cite{yagi+2013}, \cite{gavazzi+2018}\\
  GMP4437 & 10.4 [\cite{salim+2016}] & 0 [\cite{mahajan+2010}] & \cite{roberts+2020}, \cite{salim+2018}, \cite{salim+2016}\\
  GMP4463 & 9.3 [\cite{salim+2016}] & 0 [\cite{sdss}] & \cite{roberts+2020}, \cite{salim+2018}, \cite{salim+2016}\\
  GMP4281 & 9.7 [\cite{salim+2016}] & 0 [\cite{mahajan+2010}] & \cite{roberts+2020}, \cite{salim+2018}, \cite{salim+2016}\\
  GMP4236 & 8.4 [\cite{salim+2016}] & 0 [\cite{mahajan+2010}] & \cite{roberts+2020}, \cite{salim+2018}, \cite{salim+2016}\\
  NGC4853 & 10.8  [\cite{salim+2016}] & 2 [\cite{mahajan+2010}] & \cite{yagi+2010}, \cite{salim+2018}, \cite{salim+2016}, \cite{chen+2020}, \cite{gavazzi+2018}\\
  GMP4159 & 9.8 [\cite{salim+2016}] & 0 [\cite{mahajan+2010}] & \cite{roberts+2020}, \cite{salim+2018}, \cite{salim+2016}\\
  GMP4135 & 9.8 [\cite{salim+2016}] & 0 [\cite{mahajan+2010}] & \cite{roberts+2020}, \cite{salim+2018}, \cite{salim+2016}\\
  GMP4106 & 9.1 [\cite{salim+2016}] & 0 [\cite{sdss}] & \cite{roberts+2020}, \cite{salim+2018}, \cite{salim+2016}\\
  IC3949 & 10.6 [\cite{salim+2016}] & 2 [\cite{mahajan+2010}] & \cite{yagi+2010}, \cite{chen+2020}, \cite{gavazzi+2018}, \cite{salim+2018}, \cite{salim+2016}\\
  NGC4858 & 10.2 [\cite{salim+2016}] & 0 [\cite{sdss}] & \cite{chen+2020}, \cite{yagi+2010}, \cite{salim+2018}, \cite{roberts+2020}, \cite{gavazzi+2018}, \cite{salim+2018}, \cite{salim+2016}\\
  Mrk58 & 9.8 [\cite{salim+2016}] & 0 [\cite{sdss}] &\cite{chen+2020}, \cite{yagi+2010}, \cite{roberts+2020}, \cite{gavazzi+2018}, \cite{salim+2018}, \cite{salim+2016} \\
  GMP3618 & 10.1 [\cite{salim+2016}] & 6 [\cite{mahajan+2010}] & \cite{roberts+2020}, \cite{salim+2018}, \cite{salim+2016}\\
  GMP3271 & 9.1 [\cite{salim+2016}] & 0 [\cite{mahajan+2010}] & \cite{yagi+2010}, \cite{chen+2020}, \cite{roberts+2020}, \cite{gavazzi+2018}, \cite{salim+2018}, \cite{salim+2016}\\
  GMP3253 & 9.4 [\cite{salim+2016}] & 0 [\cite{mahajan+2010}] & \cite{roberts+2020}, \cite{salim+2018}, \cite{salim+2016}\\
  GMP3143 & 8.9 [\cite{salim+2016}] & 0 [\cite{mahajan+2010}] & \cite{roberts+2020}, \cite{salim+2018}, \cite{salim+2016}\\
  GMP3071 & 9.3 [\cite{salim+2016}] & 0 [\cite{sdss}] & \cite{chen+2020}, \cite{roberts+2020}, \cite{yagi+2010}, \cite{yoshida+2012}\\
  SDSSJ130006.15+281507.8 & 8.8 [\cite{salim+2016}] & 0 [\cite{sdss}] & \cite{roberts+2020}, \cite{salim+2018}, \cite{salim+2016}\\
  D100 & 9.3 [\cite{salim+2016}] & 2 [\cite{mahajan+2010}] & \cite{chen+2020}, \cite{yagi+2010}, \cite{smith+2010}, \cite{jachym+2017}, \cite{roberts+2020}, \cite{cramer+2019}, \cite{salim+2018}, \cite{salim+2016}\\
  GMP2625 & 9.1 [\cite{salim+2016}] & 0 [\cite{sdss}] & \cite{roberts+2020}, \cite{salim+2018}, \cite{salim+2016}\\
  GMP2599 & 9.9 [\cite{salim+2016}] & 4 [\cite{nucita+2017}, \cite{birchall+2020}] & \cite{chen+2020}, \cite{smith+2010}, \cite{roberts+2020}, \cite{salim+2018}, \cite{salim+2016}\\
  IC4040 & 10.3 [\cite{salim+2016}] & 5 [\cite{best+2012}] & \cite{chen+2020}, \cite{smith+2010}, \cite{yagi+2010}, \cite{yoshida+2012}, \cite{roberts+2020}, \cite{salim+2018}, \cite{salim+2016}\\
  NGC4911 & 11.3 [\cite{salim+2016}] & 2 [\cite{mahajan+2010}] & \cite{chen+2020}, \cite{yagi+2010}, \cite{roberts+2020}, \cite{salim+2018}, \cite{salim+2016} \\
  GMP2073 & 10.4 [\cite{salim+2016}] & 0 [\cite{sdss}] & \cite{roberts+2020}, \cite{salim+2018}, \cite{salim+2016}\\
  NGC4921 & 11.0 [\cite{salim+2016}] & 6 [\cite{mahajan+2010}] & \cite{kenney+2015}, \cite{chen+2020}, \cite{salim+2018}, \cite{salim+2016}\\
  GMP1616 & 10.3 [\cite{salim+2016}] & 0 [\cite{mahajan+2010}] & \cite{roberts+2020}, \cite{salim+2018}, \cite{salim+2016}\\
  GMP1582 & 8.7 [\cite{salim+2016}] & 0 [\cite{sdss}] & \cite{roberts+2020}, \cite{salim+2018}, \cite{salim+2016}\\
  GMP713 & 9.0 [\cite{salim+2016}] & 0 [\cite{mahajan+2010}] & \cite{roberts+2020}, \cite{salim+2018}, \cite{salim+2016}\\
  GMP672 & 8.7 [\cite{salim+2016}] & 0 [\cite{sdss}] & \cite{roberts+2020}, \cite{salim+2018}, \cite{salim+2016}\\
  GMP597 & 8.7 [\cite{salim+2016}] & 0 [\cite{sdss}] & \cite{roberts+2020}, \cite{salim+2018}, \cite{salim+2016}\\
  GMP522 & 9.9 [\cite{salim+2016}] & 0 [\cite{sdss}] & \cite{roberts+2020}, \cite{salim+2018}, \cite{salim+2016}\\
  GMP455 & 9.4 [\cite{salim+2016}] & 0 [\cite{mahajan+2010}] & \cite{roberts+2020}, \cite{salim+2018}, \cite{salim+2016}\\
  GMP406 & 9.0 [\cite{salim+2016}] & 0 [\cite{sdss}] & \cite{roberts+2020}, \cite{salim+2018}, \cite{salim+2016}\\
  GMP223 & 8.5 [\cite{salim+2016}] & 0 [\cite{sdss}] & \cite{roberts+2020}, \cite{salim+2018}, \cite{salim+2016}\\
  SDSSJ130545.34+285216.8 & 9.2 [\cite{salim+2016}] & 0 [\cite{sdss}] & \cite{roberts+2020}, \cite{salim+2018}, \cite{salim+2016}\\
  SDSSJ130553.48+280644.7 & 10.1 [\cite{salim+2016}] & 0 [\cite{sdss}] & \cite{roberts+2020}, \cite{salim+2018}, \cite{salim+2016}\\
  JO147 & 10.8 [\cite{merluzzi+2010}] & 3  & \cite{merluzzi+2013}, \cite{merluzzi+2016}, \cite{gullieuszik+2020}\\
  SOS90630 & 10.0 [\cite{merluzzi+2016}] & 0 [\cite{merluzzi+2016}] & \cite{merluzzi+2016}\\
  SOS61086 & 9.6 [\cite{merluzzi+2016}] & 0 [\cite{merluzzi+2016}] & \cite{merluzzi+2016}\\
  d4 & 9.8 [\cite{ebeling+2019}] & 0 [\cite{ebeling+2019}] & \cite{ebeling+2019}\\
  A1758N\_JFG1 & 10.9 [\cite{ebeling+2019}] & 0 [\cite{kalita+2019}] & \cite{ebeling+2019}\\
  d5 & 10.1 [\cite{ebeling+2019}] & 0 [\cite{ebeling+2019}] & \cite{ebeling+2019}\\
  d7 & 9.8 [\cite{ebeling+2019}] & 4 [\cite{ebeling+2019}] & \cite{ebeling+2019}\\
  d6 & 8.9 [\cite{ebeling+2019}] & 0 [\cite{ebeling+2019}] & \cite{ebeling+2019}\\
  d3 & 8.4 [\cite{ebeling+2019}] & 0 [\cite{ebeling+2019}] & \cite{ebeling+2019}\\
  d1 & 8.5 [\cite{ebeling+2019}] & 0 [\cite{ebeling+2019}] & \cite{ebeling+2019}\\
  d2 & 9.5 [\cite{ebeling+2019}] & 0 [\cite{ebeling+2019}] & \cite{ebeling+2019}\\
  C153 & 10.4 (*) & 3 [\cite{owen+2006}] & \cite{owen+2006}\\
  ESO137-001 & 9.6 [\cite{sun+2010}] & 0 [\cite{fossati+2016}] & \cite{sun+2010}, \cite{sun+2006}, \cite{sun+2007}, \cite{sun+2010}, \cite{sivandam+2010}, \cite{jachym+2019}, \cite{jachym+2014}, \cite{fossati+2016}, \cite{fumagalli+2014}, \cite{zhang+2013}\\
  ESO137-002 & 10.3 [\cite{sun+2010}] & 4 [\cite{sun+2010}] & \cite{zhang+2013}, \cite{ruszkowski+2014}, \cite{sun+2007}\\
  Sausage9 & 10.6 [\cite{sobral+2015}] & 2 [\cite{sobral+2015}] & \cite{stroe+2020}\\
  Sausage7 & 10.5 [\cite{sobral+2015}] & 0 [\cite{sobral+2015}] & \cite{stroe+2020}\\
  Sausage6 & 9.4 [\cite{sobral+2015}] & 0 [\cite{sobral+2015}] & \cite{stroe+2020}\\
  Sausage8 & 10.6 [\cite{sobral+2015}] & 2 [\cite{sobral+2015}] & \cite{stroe+2020}\\
  Sausage5 & 9.6 [\cite{sobral+2015}] & 0 [\cite{sobral+2015})] & \cite{stroe+2020}\\
  235144-260358 & 9.3 (*) & 0 [\cite{cortese+2007}] &\cite{cortese+2007}\\
\hline
\enddata
\end{deluxetable*}

\begin{table}
\begin{center}
\caption{AGN fractions for the LIT-RPS sample, considering galaxies of different mass ranges and characterized by different Jstage$_{gen}$. Errors on fractions are binomial. Values outside/in brackets are the fractions computed ignoring/considering the 4 galaxies classified as AGN based on radio and X data.
    \label{fagn_litt}}
\scriptsize
\begin{tabular}{cccc}
\hline\hline
{$N_{AGN}/N_{TOT}$} &
{$f_{AGN}$} &
{Jstage$_{gen}$} &
{$\log (M_\ast/M_\odot)$} \\
    \hline
    20/82  (24/82) & $0.24_{- 0.04}^{+0.05}$ ($0.29_{-0.05}^{+0.05}$) & all  & all \\ 
    12/15 (12/15) & $0.80_{-0.12}^{+0.08}$ ($0.80_{-0.12}^{+0.08}$) & all & $\geq$ 10.5\\
    4/15 (5/15) & $0.27_{-0.10}^{+0.13}$ ($0.33_{- 0.11}^{+0.13}$) & =0 & all\\
    2/10 (2/10) & $0.20_{-0.10}^{+0.15}$ ($0.20_{-0.10}^{+0.15}$) & =0.5 & all\\
    3/18 (4/18) & $0.17_{-0.07}^{+0.11}$ ($0.22_{-0.08}^{+0.11}$) & =1 & all\\
    8/30 (10/30) & $0.27_{-0.07}^{+0.09}$ ($0.33_{-0.08}^{+0.09}$) & =2 & all\\
    \hline
    \end{tabular}
\end{center}
\end{table}

\section{RESULTS II: Is the AGN fraction among ram-pressure stripped galaxies higher than in non-ram-pressure stripped galaxies?}

\begin{figure}
    \centering
    \includegraphics[scale=0.3]{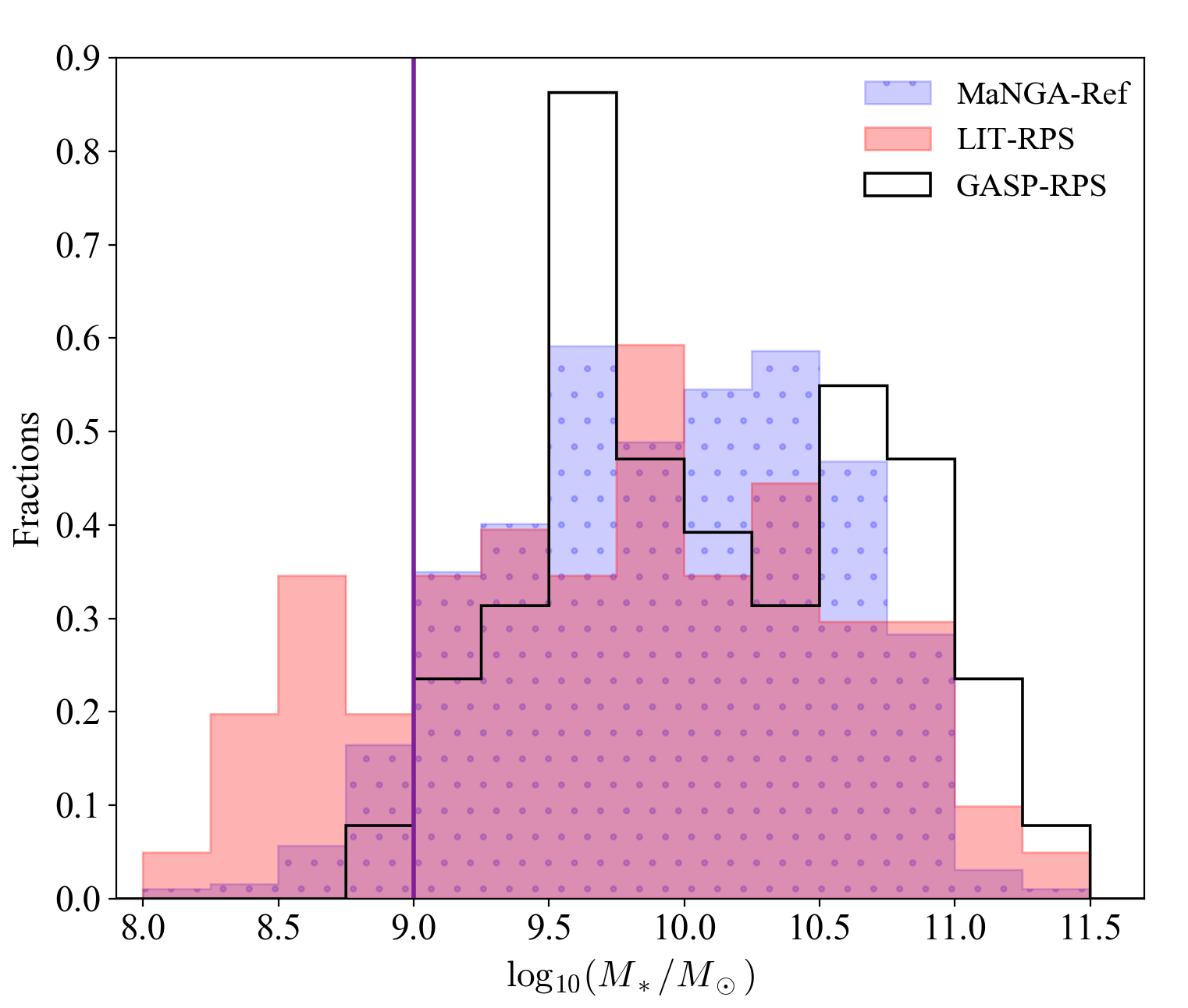}
    \caption{Normalized stellar mass distributions of the GASP (black histogram), MaNGA (purple-dotted histogram) and literature sample (light red histogram). For the Montecarlo, we have selected galaxies above the vertical-dotted line, i.e. with masses $\log(M_\ast/M_\odot) > 9$. \label{comparison}}
\end{figure}

\begin{table*}
\begin{center}
\caption{AGN fractions and binomial errorbars for the GASP-RPS, LIT-RPS and ALL-RPS samples in two different mass bins. For the LIT-RPS and ALL-RPS sample, values in parenthesis are obtained considering also galaxies identified as AGN on the basis of X-ray or radio data. 
\label{gasp_litt}}
\scriptsize
\begin{tabular}{ccccccccc}
\hline\hline
&
\multicolumn{2}{c}{GASP-RPS} &&
\multicolumn{2}{c}{LIT-RPS} &&
\multicolumn{2}{c}{ALL-RPS} \\
\cline{2-3}  \cline{5-6}  \cline{8-9}  
{$\log (M_\ast/M_\odot)$} &
{$ N_{AGN}/N_{TOT}$} &
{$\rm f_{AGN}$} & &
{$ N_{AGN}/N_{TOT}$} &
{$\rm f_{AGN}$} & &
{$ N_{AGN}/N_{TOT}$} &
{$\rm f_{AGN}$} \\
\hline
$\geq 9.0$ & 12/50 & $0.24^{+0.07}_{-0.06}$ && 19/65(23/65) & $0.29_{-0.05}^{+0.06} (0.35_{-0.06}^{+0.06})$ && 31/115(35/115) & $0.27_{-0.04}^{+0.04}$ ($0.30_{-0.04}^{+0.04}$) \\
$ \geq 10.0$  & 12/25 & $0.46^{+0.10}_{-0.09}$ &&  17/31(19/31) & $0.55_{-0.09}^{+0.09}$ ($0.61_{-0.09}^{+0.08}$) && 29/57(31/57) & $0.51_{-0.07}^{+0.07} (0.54_{-0.07}^{+0.07})$ \\
\hline
\end{tabular}
\end{center}
\end{table*}

\begin{figure}
\begin{center}
\includegraphics[scale=0.6]{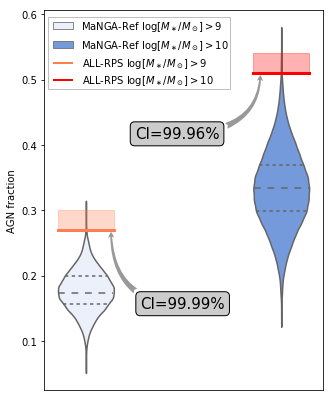}
\end{center}
\caption{Comparison of the AGN fraction in the different samples.  Red and orange  lines refer to the ALL-RPS sample: the AGN fraction for galaxies with $M_\ast>10^9 M_\odot$ is shown by the thick orange line, that for galaxies with $M_\ast>10^{10} M_\odot$ by the thick red line. The matched shaded areas indicate by how much fractions change if we consider also the AGN classified on the basis of X-ray or radio data (see text for details). Blue and light blue violin plots refer to the MaNGA-Ref sample, for the two mass bins as indicated in the labels. They show the probability density of the bootstrap random extractions mass-matched to the ALL-RPS sample, at different AGN fraction values, smoothed by a kernel density estimator. Grey horizontal dashed and dotted lines represent median values and  25\% and 75\% percentiles of the AGN fraction, respectively. Values of the pivotal confidence intervals of the bootstrap distribution are also reported: the mass-matched MaNGA fractions are lower than the ALL-RPS fractions at the 99.99\% confidence level  
 for galaxies with $M \geq 10^9 \, M_{\odot}$ and at the  99.96\% level for $M \geq 10^{10} \, M_{\odot}$.
\label{fig:AGNfrac}}
\end{figure}

In the previous section we have quantified the incidence of AGN in ram pressure stripped galaxies. 
We have seen that they represent 24\% 
of the overall, both for GASP-RPS and LIT-RPS samples.
In the following we will always exclude masses $< 10^9 \, M_{\odot}$, in all samples. Table \ref{gasp_litt} presents the AGN fractions in GASP-RPS and LIT-RPS separately for stellar masses $\geq 10^{9} M_{\odot}$ and $\geq10^{10} M_{\odot}$.\footnote{From now on we exclude from this analysis ID345 at z=0.73, considered a redshift outlier.} The fractions in the LIT-RPS sample are always higher than in GASP-RPS (0.29 vs 0.24 and 0.55 vs 0.46, respectively for the two mass bins), but are compatible within the binomial errors. Including also the 4 X-ray/radio AGN in the literature, fractions are slightly higher. We note that also the stellar mass distributions of the two samples are similar (Fig.\ref{comparison}), and indeed a Kolmogorov-Smirnov (KS) test cannot exclude that they are drawn from the same parent distribution.

It is therefore appropriate to join the two RPS samples, to obtain the largest possible statistics\footnote{Note that the galaxy JO147 appears in both samples, from now on we will just consider it once.}, and derive the total AGN fractions: $0.27^{+0.04}_{-0.04}$ at masses $\geq 10^9 \, M_{\odot}$ and $0.51^{+0.07}_{-0.07}$ for masses $\geq 10^{10} \, M_{\odot}$. These fractions are high, but less extreme than the fraction that would have been inferred from the \citetalias{poggianti+2017b} results, where 6/7 galaxies were AGN with a corresponding fraction of 0.86$^{+0.18}_{-0.09}$. This is due to the fact that the 2017 sample was composed of all massive Jstage=2 galaxies, and as we have seen in the previous sections these are the most favorable conditions for AGN activity in RPS galaxies.

We now aim at establishing whether the AGN frequency is connected to RPS and therefore we compare the measured fractions to those obtained exploiting the  MaNGA-Ref sample, used as representative of non-ram pressure stripped field galaxies. 

As for the other samples, also in MaNGA-Ref  
AGN are located preferentially among the most massive galaxies (right panel of Figure \ref{massdistr}). 
The AGN fraction is
$0.15^{+0.01}_{-0.01}$\footnote{Previous MaNGA works (e.g. \citealt{sanchez+2018}) have found a significantly lower AGN incidence. However previous analysis has not applied any cut in SSFR as we do, and have adopted much more stringent definitions (emission line ratios above the Kewley demarcation lines considering all the BPT diagrams simultaneously and H$\alpha$ equivalent width $>1.5$\AA{}{} in the central regions) for AGN, therefore results are not directly comparable.} above  $\log(M_\ast/M_{\odot}) \geq 9.0$
and $0.28^{+0.02}_{-0.02}$ for $\log(M_\ast/M_{\odot}) \geq 10.0 $.

A KS test excludes that the MaNGA-Ref mass distribution is drawn from the same parent distribution of the GASP+LIT sample (Fig. \ref{comparison}). Since the probability to find an AGN increases with galaxy mass, to properly compare the fractions obtained from MaNGA-Ref and ALL-RPS we need to control for the different mass distributions.
We perform a bootstrap random extraction of the MaNGA sample to create
10000 subsamples with the same mass distribution of the ALL-RPS sample matching the number of ALL-RPS galaxies in bins of 0.3 dex in stellar mass.
For each of the extracted samples we compute the AGN fraction $f_{AGN}$. We repeat the random extraction considering separately two stellar mass ranges, $\rm M_{\ast} \geq 10^{9} \, M_\odot$ and  $\rm M_{\ast} \geq 10^{10} M_\odot$. Violin plots with the $f_{AGN}$  distributions for the two mass ranges, their medians and the 25th and 75th percentiles are shown in Fig.\ref{fig:AGNfrac}. We find that the median $f_{AGN}$ of the 10000 realizations of mass-matched MaNGA galaxies are $f_{AGN}=$0.18 for $M \geq 10^9 \, M_{\odot}$ and 0.35 for $M \geq 10^{10} \, M_{\odot}$. 
These values are lower than the corresponding values in the ALL-RPS sample, which are 0.27 and 0.51, respectively.
In order to assess the significance of the difference between the RPS and non-RPS samples, we compute the pivotal confidence intervals of the bootstrap distribution and find that the mass-matched MaNGA fractions are lower than the ALL-RPS fractions at the 99.99\% confidence level for galaxies with $M \geq 10^9 \, M_{\odot}$  ($>$ 99.99\% if we include the 4 radio/X-ray AGNs), and at the 99.96\% level for $M \geq 10^{10} \, M_{\odot}$ ($>$ 99.99\% if the 4 radio-X-ray AGNs are included) (see Fig.\ref{fig:AGNfrac}).
 
Since the three samples considered span slightly different redshift ranges, we performed the bootstrap random extractions also limiting all samples to  $z \leq 0.075$ (the GASP redshift limit). Results remained unchanged as fractions are affected only at the 1\% level at most.
Finally, we also tried comparing mass-matched MaNGA samples separately with GASP-RPS and LIT-RPS. Though clearly the statistics decreases, we still find high probabilities that the mass-matched MaNGA sample has lower AGN fractions than the RPS samples (81.4\% and 95.4\% for GASP-RPS in the two galaxy mass ranges, and $> 99.99$\% for both LIT-RPS samples).

From our analysis the incidence of AGN activity among ram pressure stripped galaxies is significantly higher than that in the MaNGA field control sample.  A ram pressure stripped galaxy has a 1.5 times higher probability to host an AGN than a similar non-ram pressure stripped galaxy. This effect is not driven by different stellar mass distributions and points to a connection between RPS and AGN activity.

A larger (of the order of hundreds), homogeneous sample of ram-pressure stripped galaxies with integral-field spectroscopy would be needed to place these results on more solid ground. Since this is currently unavailable, the analysis presented here collects the best available datasets for addressing the question of the AGN-RPS connection. There are however several caveats which are worth stressing.

First of all, the AGN fraction depends strongly on the criteria adopted when using the BPT diagram.
In this paper, we are including both LINER-like and Seyfert AGN, in order to capture also low luminosity AGN. This is done in all samples considered in a similar manner, so it should not affect the relative incidence and the main conclusions of this work, but the pure AGN fractions will strongly depend on the initial choice.

Second, although great care has been taken to ensure the most homogeneous analysis possible, the datasets are clearly dis-homogeneous.  Even GASP and MaNGA, that are both based on integral-field data for every galaxy, have been observed with different instruments, thus have different resolutions, spaxel size etc, and span a slightly different redshift range (see above for invariance of AGN fractions with the redshift interval adopted). The literature sample, obviously, is in itself very heterogeneous, with the spectroscopic information coming from many different sources. The results shown in this paper should therefore be taken with caution, and revisited once large homogeneous samples will become available.

Third, in principle it is possible that the high AGN fraction we observe in RPS galaxies is not a consequence of RPS itself. If the AGN incidence in cluster star-forming galaxies was higher in general than in similar galaxies in the field, the differences with respect to MaNGA would go in the same direction of what we observe. However, as mentioned also above, the AGN fraction in the GASP non-ram-pressure stripped sample is small (2/49) \citep{vulcani+2021}. This sample is composed both of cluster and field ``undisturbed" galaxies. If we consider only the GASP cluster control sample (star-forming and late-types), there are no AGN (B. Poggianti et al. in prep). So, this caveat is unlikely to be responsible for our results.

Finally, we note that we are not studying the global AGN fraction in clusters, but the occurrence of AGN activity in a very specific class of cluster galaxies: those with clear signs of ram-pressure stripping, which are all star-forming and late-type galaxies and thus represent a small fraction of the total cluster galaxy population that are dominated by early-type galaxies. Therefore, our results cannot be used to infer the total AGN fraction in clusters and not necessarily show similar trends.

\section{Summary}

In this paper we have investigated the occurrence of AGN activity in  ram-pressure stripped galaxies in local clusters, comparing it with the AGN frequency in a control sample of field galaxies. In all cases, we rely on BPT diagnostic diagrams based on the [NII] line. All the galaxies analyzed in this paper are star-forming and morphologically late-type galaxies.

First, we assembled two samples of ram pressure stripped galaxies. We have used the MUSE data of 51 galaxies observed in the context of the GASP survey (GASP-RPS) finding a Seyfert2 and 4 LINER-like AGN hosts previously unknown, in addition to the 7 galaxies already discussed in \citetalias{poggianti+2017b} and \cite{fritz+2017}. We have then conducted a search in the literature assembling a sample of 82 ram-pressure stripped galaxies for which it was possible to retrieve information on their nuclear activity (either from IFU or slit/fibre) (LIT-RPS). 

We find similar fractions of AGN in GASP and in literature ram-pressure stripped  galaxies, with the AGN incidence being slightly higher in the literature than in GASP, but consistent within the uncertainties. Overall, the AGN fraction in the total GASP-RPS+LIT-RPS sample is $0.27^{+0.04}_{-0.04}$ at masses $M_* \geq 10^9 \, M_{\odot}$ and $0.51^{+0.07}_{-0.07}$ at $M_* \geq 10^{10} \, M_{\odot}$. Thus, more than half of the $\geq 10^{10} \, M_{\odot}$ ram-pressure stripped galaxies show AGN activity.

We then compare these findings with those for a sample of galaxies drawn from the MaNGA survey and inhabiting dark matter haloes with masses $\leq 10^{13} M_{\odot}$. With this halo mass cut we ensure that rich groups and clusters are excluded, hence these galaxies are not undergoing significant RPS and this can serve as a control field sample. 
We perform a bootstrap random extraction from the MaNGA sample to create 10000 realizations with the same stellar mass distribution of the ram-pressure stripped sample.

Our two main results can be summarized as follows:

1) The great majority of galaxies hosting an AGN, in all three samples considered, are high-mass galaxies. There are just very few galaxies with an AGN at masses below $10^{10} \, M_{\odot}$  (no one below $10^{10.5} \, M_{\odot}$ in GASP). As a consequence, the AGN fractions are higher above these limits, and very low below. Another factor that could be playing a role is the ram-pressure strength or phase (Jstage): the highest AGN fractions are observed among the most strongly ram-pressure stripped galaxies with the longest tails. However, with the current samples it is hard to disentangle between mass and Jstage effects.

2) Even after matching the galaxy mass distributions, the AGN incidence in the field MaNGA sample is 
lower than in the ram-pressure stripped sample at the $\geq 99.96\%$ confidence level. Overall, a ram pressure stripped galaxy has a 1.5 times higher probability to host an AGN than a similar non-ram pressure stripped galaxy. This supports the hypothesis that ram-pressure can trigger the AGN activity.
 
\section*{Acknowledgements}
We thank the Referee for the useful suggestions that improved the presentation of the work. We warmly thank Jong-Ho Shinn from the Korea Astronomy and Space Science Institute for useful discussion regarding the statistical analysis.   GP thanks A. Werle for the helpful discussion. Based on observations collected at the European Organization for Astronomical Research in the Southern Hemisphere under ESO programme 196.B-0578. This project has received funding from the European Research Council (ERC) under the European Union's Horizon 2020 research and innovation programme (grant agreement No. 833824).  We acknowledge financial contribution from  the grant PRIN MIUR 2017 n.20173ML3WW\_001 (PI Cimatti), from the INAF main-stream funding programme (PI Vulcani) and from the agreement ASI-INAF n.2017-14- H.0 (PI A. Moretti). Y.J. acknowledges financial support from CONICYT PAI (Concurso Nacional de Inserci\'on en la Academia 2017) No. 79170132 and FONDECYT Iniciaci\'on 2018 No. 11180558. J.F. acknowledges financial support from the UNAM- DGAPA-PAPIIT IN111620 grant, México.

\bibliography{biblio}{}
\bibliographystyle{aasjournal}

\appendix

\section{Mass estimates} \label{sec:appendix_mass}
 To compute stellar masses for those galaxies for which they are missing in the literature. we use the \cite{bell_dejong+2001} relation between the mass-to light ratio of a galaxy and its color:

\begin{equation}
     \log\left(\frac{M}{L_{\lambda}}\right) = a_{\lambda} + b_{\lambda} \cdot \rm COL
     \label{bell_jong}
\end{equation}

where $L_{\lambda}$ is the luminosity in a band, indicated with $\lambda$, COL is a photometric color and $a_{\lambda}$ and $b_{\lambda}$ are coefficients depending on both $\lambda$ and COL.
For our calculations, we used the \citeauthor{bell_dejong+2001} tables for a solar metallicity Z = 0.02 and a \cite{bruzualcharlot2003} SSP model, 
converting from a \cite{salpeter1955} to a \cite{chabrier+2003} IMF subtracting a factor -0.24.
For one galaxy, 235144-260358 \citep{cortese+2007}, in order to use the formula (\ref{bell_jong}) we first converted \emph{HST} magnitudes to an UBV phometric system with the use of calibration equations for the Advanced Camera for Surveys (ACS) presented in \cite{sirianni+2005}.
We assume a typical 0.3 dex uncertainty on the computed stellar masses, which we take as bin size of the stellar mass distribution.

\end{document}